\documentclass[journal ]{new-aiaa}
\usepackage[utf8]{inputenc}
\usepackage{textcomp}

\usepackage[table,xcdraw]{xcolor}
\usepackage{empheq}
\usepackage{subfig}
\usepackage{systeme}
\usepackage{tikz}
\usepackage{graphicx}
\usepackage{amsmath}
\usepackage[version=4]{mhchem}
\usepackage{siunitx}
\usepackage{longtable,tabularx}
\title{Drone Acoustic Analysis for Predicting Psychoacoustic Annoyance via Artificial Neural Networks}

\author{Vaiuso Andrea \footnote{Research assistant, Lightweight Design, Andrea.Vaiuso@zhaw.ch}, Righi Marcello\footnote{Professor, AIAA Member, Lecturer at Federal Institute of Technology Zurich ETHZ, Marcello.Righi@zhaw.ch}, Coretti Oier \footnote{Research assistant, Lightweight Design, Oier.Coretti@zhaw.ch}, Apicella Moreno \footnote{Master Student, Lightweight Design, Moreno.Apicella@zhaw.ch}, }
\affil{ZHAW School of Engineering, Winterthur, Switzerland, 8400}

\begin{document}

\maketitle

\begin{abstract}
Unmanned Aerial Vehicles (UAVs) have become widely used in various fields and industrial applications thanks to their low operational cost, compact size and wide accessibility. However, the noise generated by drone propellers has emerged as a significant concern. This may affect the public willingness to implement these vehicles in services that require operation in proximity to residential areas. The standard approaches to address this challenge include sound pressure measurements and noise characteristic analyses. The integration of Artificial Intelligence models in recent years has further streamlined the process by enhancing complex feature detection in drone acoustics data.
This study builds upon prior research by examining the efficacy of various Deep Learning models in predicting Psychoacoustic Annoyance, an effective index for measuring perceived annoyance by human ears, based on multiple drone characteristics as input. This is accomplished by constructing a training dataset using precise measurements of various drone models with multiple microphones and analyzing flight data, maneuvers, drone physical characteristics, and perceived annoyance under realistic conditions. The aim of this research is to improve our understanding of drone noise, aid in the development of noise reduction techniques, and encourage the acceptance of drone usage on public spaces. 

\end{abstract}

\section*{Nomenclature} 

{\renewcommand\arraystretch{1.0}
\noindent\begin{longtable*}{@{}l @{\quad=\quad} l@{}}
$DL$ & Disc Loading (Pa) \\
$F$  & Fluctuation Strength (vacil) \\
$f_{mod}$ & Modulation Frequency \\
$H$  & Humidity (\% / 100) \\
$l_{arm}$  & Drone arm length (mm) \\
$l_{rot}$  & Drone's rotor radius (mm) \\
$L$  & Loudness (sone) \\
$L'$ & Specific Loudness (sone) \\
$\Delta L$ & Perceived Modulation Depth \\
$N_{rot}$  & Number of rotors \\
$P$  & Pressure (hPa) \\
$PA$  & Psychoacoustic Annoyance \\
$R$  & Roughness (asper) \\
$S$  & Sharpness (acum) \\
$T$  & Time (s) \\
$W$  & Weight (Kg) \\
$Wi_{avg}$  & Wind average velocity (m/s) \\
$Wi_{max}$  & Wind max velocity (m/s) \\
\end{longtable*}}

\section{Introduction}
Unmanned Aerial Vehicles (UAVs), commonly known as \textit{"drones"}, have emerged as versatile tools with a wide range of applications in various sectors, including transportation, communication, agriculture, disaster management, and environmental conservation~\cite{floreano2015science}. Moreover, in industrial applications, UAVs perform vital monitoring tasks, such as Structural Health Monitoring (SHM) for hard-to-reach locations~\cite{catt2019development}. Moreover, in the field of delivery services, UAVs are proposed as a more environmentally friendly alternative, potentially reducing greenhouse gas emissions and other environmental impacts~\cite{goodchild2018delivery}.

Despite their numerous advantages, the rapid adoption of drones for these applications faces challenges related to public acceptance. Safety concerns and disruptions to personal environments, particularly noise pollution, are among the primary issues. Recent studies have demonstrated that people find drone noise more disturbing than noise from conventional ground vehicles and full-size aircraft~\cite{cabell2016measured}. This noise aversion extends to wildlife as well~\cite{raya2017small}. This increased annoyance can be attributed to the specific aerodynamic characteristics of drones, such as blade and body design, rotor spacing, blade passing frequency (BPS), and all the various interactions between the rotors and the drone structure. Small UAVs using all-electric propulsion are expected to be quieter than traditional aircraft, but as they operate closer to the public, noise annoyance becomes a concern. Legislators and stakeholders require efficient tools to predict and assess drone noise levels without relying solely on field testing, which can be inconvenient and lacks preemptive assessment capabilities. This has led to the development of effective tools for predicting drone noise exposure in complex urban and extra-urban settings.

\subsection{Featured works and goal}
Efforts have been made to measure and understand the noise generated by drones, including studies that involve sound pressure measurements and the characterization of noise emission patterns. These studies have contributed to our knowledge of drone noise in various scenarios and have laid the foundation for predicting and managing noise exposure. Sinibaldi et al.~\cite{sinibaldi2013experimental} explored ways to optimize rotor design to reduce noise impact. However, this study primarily focused on propeller noise, without considering the aerodynamic interactions within a full drone structure. Cabell, et al.~\cite{grosveld2016measured} documented sound pressure levels throughout drone flights, offering an angle-dependent emission model. Alexander et. al.~\cite{alexander2019flyover,alexander2019predicting} meticulously documented sound pressure measurements during drone operations, whether hovering or navigating at low speeds above grassy terrains. Their studies involved the strategic placement of microphones on acoustically rigid ground plates. In parallel, Besnea~\cite{besnea2020acoustic} harnessed the power of microphone arrays for pinpointing the sources of acoustic emissions, yet encountered hurdles in discerning source strength.  Furthermore, recent research by Gallo et al.~\cite{gallo2022annoyance} aims to correlate several Sound Quality Metrics (SQM) with perceived annoyance (PA) by analyzing drone flight maneuvers. By using multiple microphones in both an anechoic chamber and real-world settings, this work builds upon previous studies by Torija et al.~\cite{torija2019psychoacoustic}, which compared SQM and PA between drones, airplanes, and ground vehicles, highlighting the specific frequency characteristics that make drone noise more bothersome. 

Artificial Intelligence (AI), and in particular Deep Neural Networks (DNN), has been used in recent years to predict and analyze complex data related to drone sound characteristics. Its remarkable capabilities have been showcased in various tasks, including fault diagnosis~\cite{iannace2019fault,ghazali2022vibration} and drone presence detection~\cite{anwar2019machine}. The multifaceted nature of drone operations demands a sophisticated understanding of their interactions with the environment, and AI algorithms excel in discerning intricate patterns within vast datasets, enabling enhanced insights into the dynamics of drone behavior. 

Our study is based on the approach in~\cite{gallo2022annoyance} and~\cite{torija2019psychoacoustic}, extending previous researches by exploring correlations between different drone's specifications, flight data, and maneuvers performed, and attempting to predict noise annoyance factors under realistic conditions. In our study, Psychoacoustic Annoyance (PA) is assessed by using four different AI-based and classic regression models, whose performance is compared and evaluated. The objective of this study is to demonstrate the applicability of this type of models in solving problems involving acoustics and psychoacoustics. This approach aims to improve our understanding of drone noise and its impacts using an AI-based model, facilitating the development of noise mitigation strategies and improving public acceptance. The section \ref{sec:methods} of this paper aims to examine the methods used to collect and process audio data. Subsequently, the section \ref{sec:analysis} will deal with the analysis of the methodology used for collecting data, the calculation of Sound Quality Metrics (SQM) and Perceived Annoyance (PA), and the comparison of these metrics with variations in the drones, microphones, and maneuvers used. A correlation map is presented and analyzed to better understand which variables influence each other. Finally, section \ref{sec:model}, we examine the methodology used for the implementation of four different regression models: Linear, Support Vector Machine (SVM), Deep Neural Network (DNN), and Convolutional Neural Network (CNN). Our analysis shows that DNN is the most effective model for predicting annoyance based on the collected data.

\section{Data Collection Methodology}\label{sec:methods}
\begin{figure}[]
    \centering
    \includegraphics[width=1\linewidth]{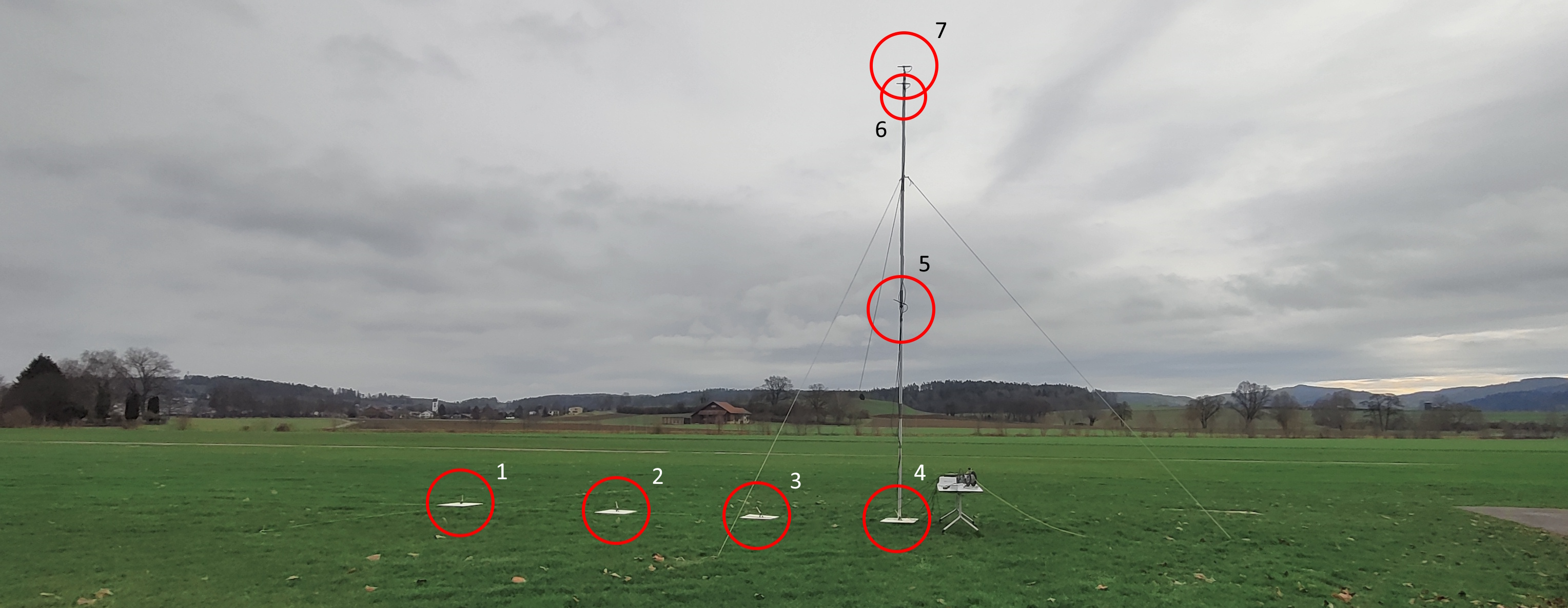}
    \caption{Final microphone setup. Each number represent the microphone ID.}
    \label{fig:setup}
\end{figure}

The process of measuring drones is far from being standardized. This lack of uniformity becomes even more pronounced when we focus on the acoustic measurement of drones during maneuvers. The ISO 5305 standard~\cite{ISO5305} has served as a valuable reference during the creation of this methodology for measuring drone noise. The primary goal in adhering to these initial recommendations from ISO is to ensure that the results obtained can be as comparable as possible. Consequently, it becomes necessary to establish a custom-defined procedure to conduct these measurements. The conceptual definition of the Methodology involve four main aspects: The selection of the drones, the definition of manoeuvres, the design of the microphone setup (or microphone array) and the definition of a measuring procedure.

All the measurements were carried out in the Riedbachweg flight area in Winterthur, Switzerland, a quiet location where drone sounds could be recorded with relatively minimal background noise. The choice to record in an open space, rather than an anechoic chamber, was deliberate to facilitate the examination of emission, propagation, and human perception of the noise in a setting that closely resembles real-world conditions. This approach takes into account all the attenuation effects and anomalies caused by atmospheric conditions, including factors such as wind, thermal inversion, and turbulence. Table \ref{tab:drone_data} and \ref{tab:setup_data} include technical details concerning the setup configurations, drone specifications, and environmental conditions during the data collection phase.

\subsection{Drones selection and maneuver definition}
In this study, we examined four drones, selected based on a compromise between availability and diversity in terms of size and configuration. Details outlining the key characteristics of the drones used can be found in Table \ref{tab:drone_data}. This selection includes two commercially available drones, a self-assembled Quadcopter (Holybro S500), and a self-assembled Hexacopter (Tarot X6).

\begin{table}[!]
    \begin{tabular}{|c|c|c|c|c|c|c|c|c|c|c|c|}
      \hline
      \rowcolor[HTML]{EFEFEF} 
      Drone & $l_{arm}$ & $l_{rot}$ & $N_{rot}$ & $W$ & Wind & $Wi_{avg}$ & $Wi_{max}$ & $T$ & $H$ & $P$ & Date \\
      \hline
      DJI Matrice & 452.5 & 267.5 & 4 & 6.30 & Ne-NW & 8.00 & 11.70 & 21.00 & 0.49 & 1018 & 01/06/2023 \\
      \hline
      DJI Mavic & 177.5 & 110 & 4 & 0.92 & N-NW & 7.50 & 14.30 & 22.50 & 0.52 & 1015 & 02/06/2023 \\
      \hline
      Holybro S500 & 242.5 & 127.5 & 4 & 0.78 & NE & 13.00 & 19.50 & 21.50 & 0.47 & 1018 & 07/06/2023 \\
      \hline
      Tarot X6 & 360 & 167.5 & 6 & 2.30 & NE-N & 14.00 & 22.10 & 24.50 & 0.37 & 1016 & 07/06/2023 \\
      \hline
    \end{tabular}
    \caption{Drone structural data and measurement environmental data (values explanation and dimensions available on nomenclature table)}
    \label{tab:drone_data}
\end{table}

The normative outlines a set of standard maneuvers that accurately represent the drone flight. Drone movement typically involves a sequence of vertical and horizontal displacements, ideally executed at a constant velocity. These movements, in combination with various drone-to-observer positions, have led to the definition of seven standardized reference maneuvers: Hover, Yaw, Ascent, Descent, Front Cruise (movement toward the observer), Back Cruise (movement away from the observer), and Lateral Cruise (sideways movement). The chosen velocity for these maneuvers is 3 m/s.

\subsection{Microphone setup}
It's essential that the setup must be adjusted for each drone size. While this presents practical challenges, it ensures that the results can be compared consistently across different drones, while also guaranteeing that all microphones are situated in the acoustic far-field. For each drone, a distinct setup configuration is defined, yet these setups share common characteristics, with minor adjustments in hovering altitude (Hr) and microphone spacing (d, hs, hi). The primary deviations between the proposed scheme and the ISO 5305 recommendation are as follows:
\begin{itemize}
    \item The structure incorporates seven microphones, including three of the four microphones specified in ISO 5305 ("m1," "m4," "m6"), along with the addition of four extra microphones ("m2," "m3," "m5").
    \item In place of the fourth microphone suggested by ISO 5305 to measure noise over the drone, a microphone ("m7") is positioned at a significantly lower height. This change was implemented for practical reasons, reducing the overall height of the support structure.
\end{itemize}
The primary structure for microphone placement is based on a 50x50mm Item profile and divided into four 2-meter-long segments for ease of transport. The microphones are affixed to adjustable supports and then secured to brackets fixed to the profile. This arrangement allows precise positioning of the microphones according to the predefined plan, ensuring they are oriented toward the hovering position. Microphone orientation is established prior to the structure's assembly by replicating the measurement conditions on the floor. The microphone setup is represented on Table \ref{tab:setup_data} for each drone selected.

\begin{table}
  \begin{minipage}{.25\textwidth}
    \centering
    \includegraphics[width=\textwidth]{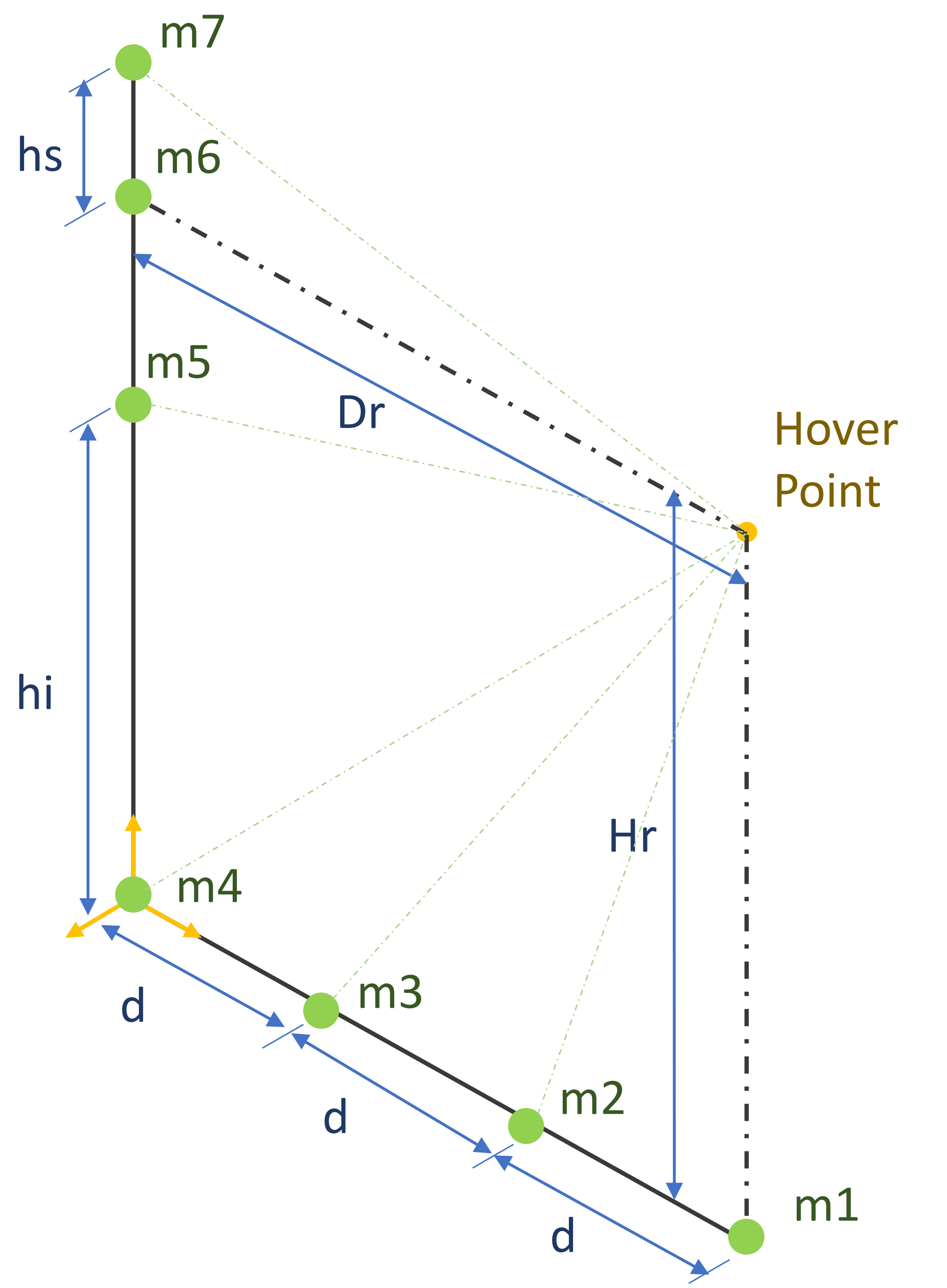}
  \end{minipage}%
  \begin{minipage}{.75\textwidth}
    \centering
    \begin{tabular}{|c|c|c|c|c|c|}
      \hline
      \rowcolor[HTML]{EFEFEF} 
      Drone        & Dr [m] & d [m] & Hr [m] & hs [m] & hi [m] \\ \hline
      DJI Matrice   & 7.755  & 2.585 & 7.755  & 0.375  & 3.878  \\ \hline
      DJI Mavic    & 6.025  & 2.008 & 6.025  & 0.375  & 3.013  \\ \hline
      Holybro S500 & 6.355  & 2.118 & 6.355  & 0.375  & 3.178  \\ \hline
      Tarot X6     & 6.985  & 2.328 & 6.985  & 0.375  & 3.493  \\ \hline
    \end{tabular}
  \end{minipage}
  \caption{Setup distances for each drone.}
  \label{tab:setup_data}
\end{table}

Behringer ECM-8000 microphones were chosen. Calibration of the microphones is also conducted at this stage. This process helps confirm that the loudness was correctly measured before the drone's flight, and it further verifies the consistency between measurements taken before and after the data collection. The Bruel \& Kjaer 4230 calibrator was employed for this precise purpose. An Audio Interface was needed that could accept 7 inputs simultaneously; in this case the selected device was the MOTU 8 Pre USB. Once the structure is fully assembled, with the microphones in their designated positions, correctly oriented, and calibrated, it can be erected for use.

\subsection{Measuring procedure}
Acquiring the acoustic measurements from all seven inputs, as utilized in the Audio Interface system, was efficiently achieved using the Performer Lite program. However, the subsequent interpretation and analysis of this data introduced two main complexities to be taken into account.

First, the decision to amalgamate all seven distinct maneuvers into a single flight introduced a level of intricacy. While this approach considerably streamlined the measurement process, it necessitated segmenting and filtering the measurements to isolate the noise during each maneuver. To address this, videos were recorded to capture the initial moments when measurements began and the specific times at which each maneuver was executed.

Secondly, maintaining accurate positional records throughout the maneuvers was essential. As previously mentioned, the positioning accuracy of each measurement plays a key role in generating valuable acoustic data. The position data, stored by the drones themselves, required thorough filtering and analysis. To account for minor deviations in the take-off and landing positions, which are geometrically identical, a small linear correction was applied.

\section{Data analysis}\label{sec:analysis}
The data analysis was divided into several phases:
\begin{enumerate}
    \item \textbf{Data Preprocessing:} All the data was saved in a hard disk memory, and standardized with a sampling rate of 22050 Hz.
    \item \textbf{Data Alignment with Flight Logs:} The aim of this phase is to align all data collected through a synchronization process. This involved matching the RPM variation peaks observed in the available flight logs with the acoustic spectrum of the audio data. Afterwards, all signals were linear interpolated to match the sound sampling rate. This alignment was easily achieved using the clock data within the logs.
    \item \textbf{Maneuver Segmentation:} With the aid of videos, it was possible to accurately isolate the various maneuvers performed by all the drones: at the end of this phase, it became feasible to analyze, for each drone, microphone, and maneuver, both the acoustic and electrophysical propulsion data.
    \item \textbf{Frequency features analysis:} Frequency signals and spectrograms were calculated for all audio signals, and examined in paragraph \ref{subsec:spectra}. 
    \item \textbf{Calculation of Acoustic Quality Metrics (SQM):} Metrics such as "loudness," "sharpness," "roughness," and "fluctuation strength" were computed. These metrics are explained and analyzed in paragraph \ref{subsec:metrics}.
    \item \textbf{Calculation of Psychoacoustic Annoyance (PA) based on SQM:} An overall annoyance index was calculated. This calculation is explained and analyzed in paragraph \ref{subsec:pa}.
    \item \textbf{Drone acoustic features comparison:} An example of SQM signals and PA values, which were the outcomes of the preceding steps, are depicted in paragraph \ref{subsec:analyresult}. Subsequently, a comparative analysis of all the drones and microphones used in the experiment was conducted, considering their SQM and PA values.
\end{enumerate}

\subsection{Acoustic spectra and spectrograms} \label{subsec:spectra}

The Fourier transform of the signals and spectrograms is calculated for all drones, all microphones, and all maneuvers. To calculate the frequency signals and spectrograms, the Python library Librosa \cite{mcfee2015librosa} is used. Specifically, the Short-Time Fourier Transform (STFT) algorithm is employed for signal in frequency calculation, where each audio frame is windowed using a window of length 2048. Scaling is applied according to the algorithm defined in \texttt{"librosa.amplitude\_to\_db"}. For spectrograms, Matplotlib library is used with $N_{FFT}$ = 2048 and \texttt{"symlog"} as a scaling factor.

\begin{figure}[!t]
    \centering
    \includegraphics[width=\textwidth]{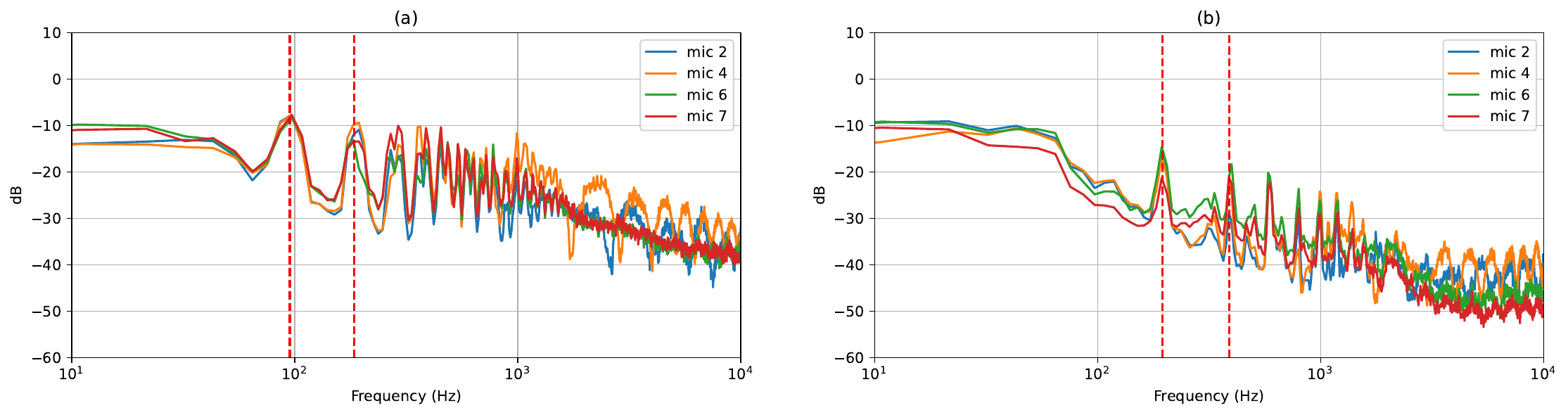}
    \caption{(a) Microphone 2, 4, 6, 7 spectrum for DJI Matrice 300 RTK during hovering. (b) Microphone 2, 4, 6, 7 spectrum for DJI Mavic 2 Enterprise during hovering.} \label{fig:freq_comp}
\end{figure}
\begin{figure}[!t]
    \centering
    \includegraphics[width=\textwidth]{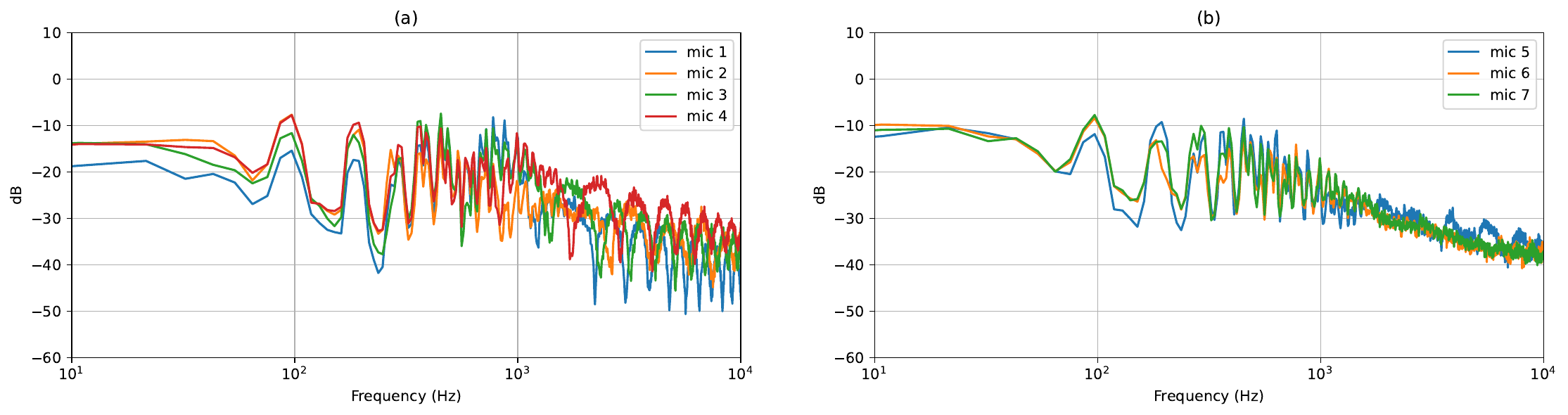}
    \caption{Comparison between (a) ground microphone spectrum versus (b) on-top microphone spectrum for DJI Matrice 300 RTK} \label{fig:freq_comp2}
\end{figure}

\begin{figure}[!t]
    \centering
    \includegraphics[width=\textwidth]{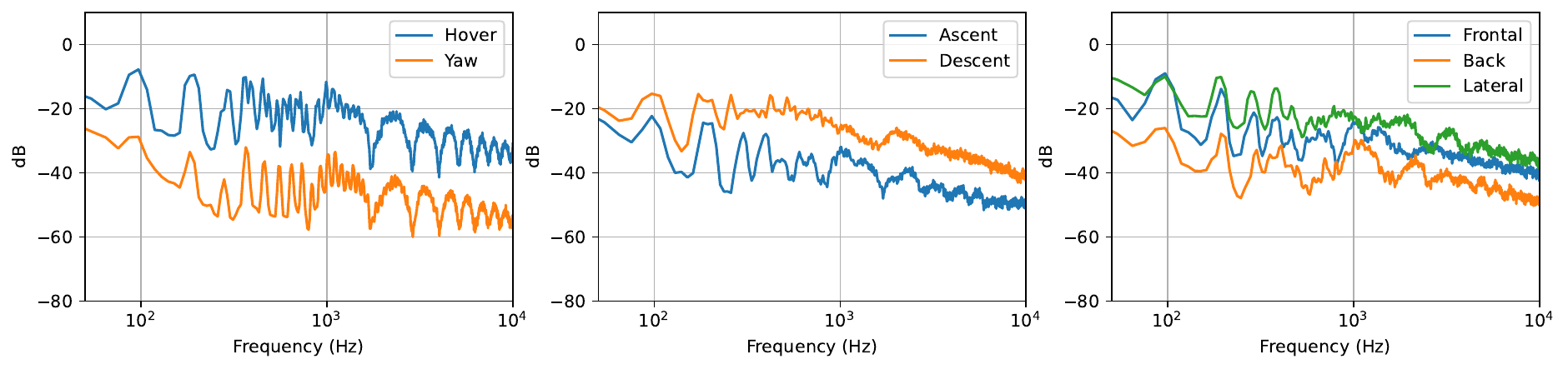}
    \caption{Fourier transformed signals for all maneuvers for DJI Matrice 300 RTK.} \label{fig:fourier_transforms}
\end{figure}

 A visual inspection immediately suggests two fundamental aspects to consider about on frequency properties. 
\begin{itemize}
    \item Each drone exhibits distinctive low-frequency peaks that are consistently captured from various observer positions. This common characteristic generally results in a high roughness value, a sound quality metric discussed in paragraph \ref{subsec:metrics}, which heavily relies on low-frequency signal components. To illustrate this, Figure \ref{fig:freq_comp} provides a comparison between two drones from our fleet during the hovering phase, recorded from all the microphones (for clarity, some signals are omitted in the figure). DJI Matrice 300 RTK (a) exhibits its initial low-frequency peaks at around 95 and 185 Hz, while DJI Mavic 2 (b) displays them at approximately 195 and 390 Hz in both input signals. Moreover, Figure \ref{fig:freq_comp2} demonstrates that, for each maneuver (in this example, only DJI Matrice 300 RTK is shown), these peaks consistently occupy the same frequencies. This observation indicates that these specific low-frequency components primarily stem from the inherent acoustic properties of the drone itself. In other words, these shared low-frequency features remain unaffected by the observer's position or the specific maneuver performed. This suggests that these low-frequency peaks are a direct consequence of the drone's distinctive structural and mechanical attributes, which consistently manifest in the audio signals recorded by all microphones across diverse scenarios.
    \item In contrast, higher frequencies seems to be affected by the drone's position relative to the observer. As depicted in Figure \ref{fig:fourier_transforms}, the spectra reveal sharper, more pronounced high peaks when recorded by ground microphones (microphones 1, 2, 3, 4), while top-mounted microphones (microphones 5, 6, 7) capture noisier and less prominent peaks within the same frequency range. This observation suggests that the sound characteristics emanating from the top of the drone generally have fewer high-frequency components. Importantly, this characteristic remains consistent across maneuvers and drone types, indicating that it is primarily associated with the drone's positioning relative to the observer, rather than the specific drone or maneuver.
\end{itemize}

\begin{figure}[!b]
    \centering
    \includegraphics[width=\textwidth]{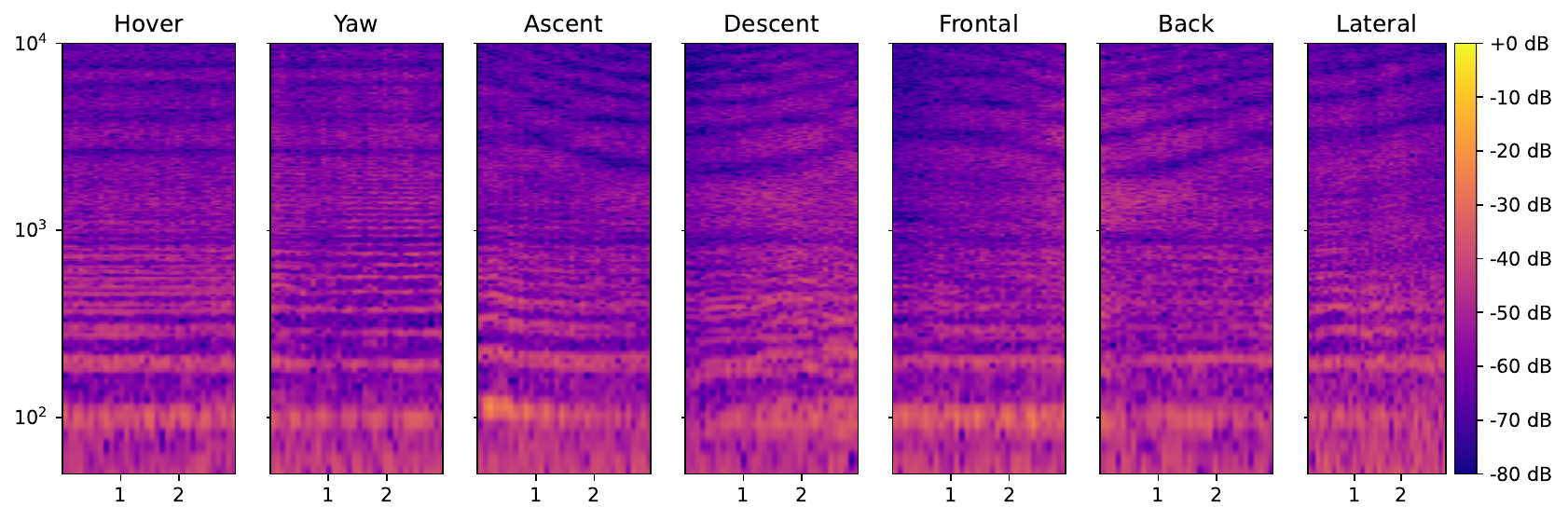}
    \caption{Spectrograms for all maneuvers for DJI Matrice 300 RTK.} \label{fig:spectrograms}
\end{figure}

Figure \ref{fig:spectrograms} shows the spectrograms of the same signals shown in Figure \ref{fig:fourier_transforms}. Again, the position of the low-frequency peaks remains almost unchanged during different maneuvers. The spectrograms provide valuable insights into the correlation between rotors speed and the variation of peaks over time. For example, while the "hover" medium-frequency signal remains quite stable, the "yaw" signal exhibits fluctuations between $10^2$ and $10^3$ Hz compared to the first one. In addition, the high frequencies give us information about the acceleration and deceleration of the drone during ascent, descent, forward, backward and lateral movement phases, showing clear and smooth curves over time.

\subsection{Sound Quality Metrics} \label{subsec:metrics}
Sound Quality Metrics (SQM) is a set of indices calculated by algorithms specific to the field of psychoacoustics~\cite{zwicker2013psychoacoustics}. SQM are employed to establish the connection between the physical attributes of sound and the subjective impression perceived by the human ear. These algorithms attempt to analyze parameters like sound pressure level, frequency, and modulation depth, and link them to human auditory perception. Sound quality encompasses several algorithms, such as ISO 532B stationary loudness, time-varying loudness, roughness, sharpness, tonality, and fluctuation strength. To compute Psychoacoustic Annoyance, calculations were performed for time-varying loudness, roughness, sharpness, and fluctuation strength.
\begin{itemize}
    \item \textbf{Loudness (sone):} Loudness is a measure for how humans perceive the volume of sound. In the definition of loudness, it is specified that 1 sone, the unit of loudness, is equivalent to a 1 kHz tone at 40dB. The algorithm for loudness computation, also  known as ISO 532B loudness or Zwicker loudness, is in compliance with ISO 532B, DIN 45631, and ISO/R 131. In particular, the algorithm for calculating time-varying loudness in compliance with DIN 45631/A. Loduness formula is defined by:
    \[
    L = \int_{z=0}^{24} L'(z) \, dz
    \]
    Where the integral is taken over all the critical-band rates, and $L'$ is referred to the specific loudness, that can be calculated from the dB level for each third octave band using a power law (see ISO BS 532/R and DIN 45631 which contains the algorithm explanation for loudness calculation.)
    \item \textbf{Roughness (asper):} Roughness is a different metric employed to assess the subjective evaluation of sound quality. It is associated with how noticeable or bothersome a sound is when perceived by the human ear. To be more precise, roughness refers to an auditory sensation linked to fluctuations in loudness occurring at frequencies too high to be individually distinguished, typically involving modulation frequencies exceeding 30 Hz. In particular, roughness formula is defined as:
    \[
    R = cal \int_{z=0}^{24} f_{mod} \, \Delta L \, dz
    \]
    Where $f_{mod}$ represents the modulation frequency, detected by peak-picking a frequency-domain representation of the acoustic loudness, $cal$ is a constant used to maintain the reference signal's roughness at a constant value of one, and $\Delta L$ is the modulation depth, that is a measure of the strength of amplitude modulation in a signal and can be calculated by finding the envelope of filtered signals and taking the difference between maximum and minimum values of the envelope. In other words, it measures how much the amplitude of the signal varies due to modulation.
    \item \textbf{Sharpness (acum):} Sharpness is a metric that captures the hearing sensation associated with frequency. It characterizes the perception of a sharp, piercing, high-frequency sound and is based on the comparison of the amount of high-frequency energy to the overall energy. The sharpness algorithm derives sharpness from the sound pressure signal waveform, the 1/3-octave band spectrum computed across the frequency range of 25 Hz to 12.5 kHz, or the specific loudness. The algorithm that computes sharpness normalizes the specific loudness spectrum by the total loudness and adjusts the spectrum based on frequency. The result, a frequency-weighted value representing specific sharpness in relation to critical band rate, is then integrated to yield the sharpness measurement. Typically, the presence of higher frequency components in the signal leads to higher sharpness readings. Sharpness formula is defined by:
    \[
    S = \frac{k}{L} \int_{z=0}^{24} L'(z) \, g(z) \, z \, dz
    \]
    Where $L$ is the loudness, $L'$ is the specific loudness in sones/Bark, the function $g(x)$ and the scaling factor $k$ depend on the specified weighting method in compilance with DIN 45692~\cite{din200945692}, Von Bismark and Aures. For our purposes, DIN 45692 method was chosen.
    \item \textbf{Fluctuation Strength (vacil):} Fluctuation strength is a metric that tries to capture the hearing sensation associated with loudness variations at low frequencies that can be individually perceived. It employs a similar approach to the "roughness versus time" analysis but is specifically focused on fluctuations with very low modulation frequencies. The fluctuation strength metric assesses the energy within 47 overlapping barks, calculates and filters the envelope of the signal in each band, measures the amplitude modulation of each envelope, and adjusts the level in each band using a frequency-dependent weighting function. The algorithm provides the result as the fluctuation strength spectrum relative to critical band rate and then integrates this spectrum to quantify the fluctuation strength. The analysis encompasses modulations between 0 to 30 Hz, with a particular emphasis on those occurring near 4 Hz. Fluctuation Strength formula is defined by:
    \[
    F = \frac{0.008 \int_{z=0}^{24} \Delta L \, dz}{{\frac{f_{mod}}{4}} + \frac{4}{f_{mod}}}
    \]
    Where, again, $f_{mod}$ represents the modulation frequency and $\Delta L$ is the perceived modulation depth
\end{itemize}
All of the metrics were calculated on Matlab software, using the Audio Toolbox package.

\subsection{Psychoacoustic Annoyance} \label{subsec:pa}
The Psychoacoustic Annoyance (PA) model, as devised by Zwicker and Fastl~\cite{zwicker2013psychoacoustics}, provides a framework for understanding the connection between several auditory sensations, including loudness, sharpness, fluctuation strength, and roughness. This model is expressed through the following mathematical formulation:
\[
    PA=N_5(1+\sqrt{w_S^2 + w_{FR}^2})
\]
Where $N_5$ is the 5th percentile of the loudness (in sones), $w_S$ and $w_{FR}$ are functions of $S$, $F$ and $R$ where:
\[
w_S = 
\begin{cases}
(S-1.75) \cdot 0.25 \log_{10} (N_5 + 10) & \text{if } S > 1.75, \\
0 & \text{if } S \leq 1.75.
\end{cases}
\]
\[
w_{FR} = \frac{2.18}{N_5^{0.4}}(0.4\cdot F + 0.6 \cdot R)
\]

The PA model offers valuable insights into how these specific psychoacoustic attributes collectively contribute to the perception of annoyance when exposed to various sound stimuli. It serves as a tool to analyze and predict the impact of sound on human annoyance, facilitating a deeper understanding of the subjective reactions to different acoustic environments.

\subsection{Results} \label{subsec:analyresult}
\begin{figure}
    \centering
    \includegraphics[width=1\linewidth]{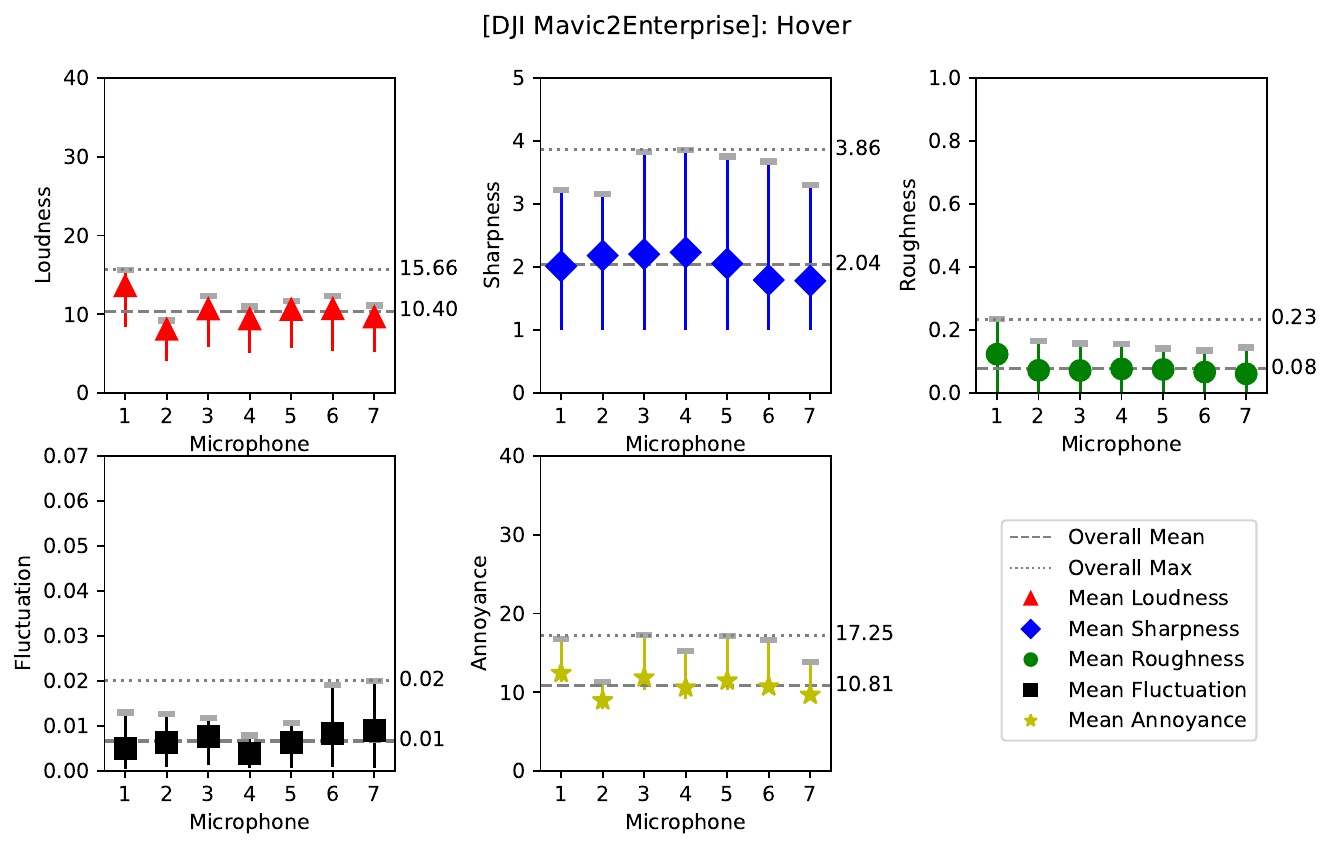}
    \caption{DJI Mavic 2 Enterprise error plot for SQM values and PA}
    \label{fig:comparative_mavic}
\end{figure}

The calculation of Sound Quality Metrics (SQM) and Psychoacoustic Annoyance (PA) was performed for each drone, maneuver, and microphone. Once again, for ease of understanding, error plots \ref{fig:comparative_matrix} and \ref{fig:comparative_mavic} only display values for the DJI Matrice and DJI Mavic 2 drones during the hovering phase, but these considerations result valid for all other drones. In the figures, markers represent the average signal values, while the segments represent the maximum peaks. From the conducted study, several observations can be made:

\begin{figure}
    \centering
    \includegraphics[width=1\linewidth]{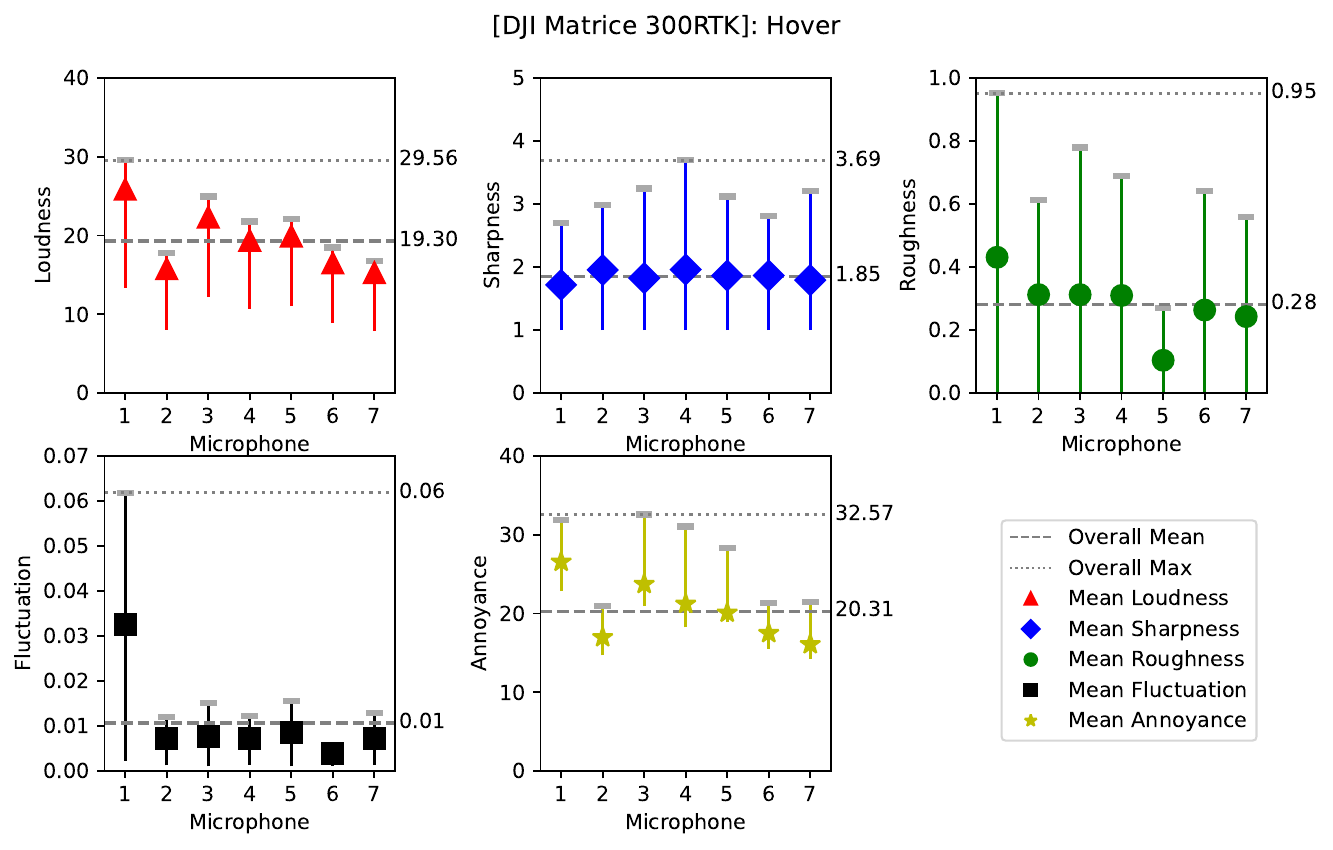}
    \caption{DJI Matrice 300 RTK error plot for SQM values and PA. Gray bars indicates maximum values.}
    \label{fig:comparative_matrix}
\end{figure}

\begin{enumerate}
    \item Overall loudness is significantly lower for the Mavic compared to the Matrice.
    \item Sharpness values remain almost constant across different microphones. The Mavic exhibits slightly higher sharpness values compared to the Matrice.
    \item Roughness values are noticeably lower for the Mavic. This observation is also consistent with the analysis in \ref{subsec:spectra}, as the low-frequency peaks of the Matrice are broader and further back in the spectrum compared to the Mavic.
    \item Fluctuation values hover around average values of about 0.01 vacil, with particularly high peaks recorded by microphone 1 during the Matrice measurement. This may be due to unexpected noise, but the average values remain relatively consistent across all drones in the experiment.
    \item Finally, annoyance appears to be quite evidently linked to loudness values. Once again, the DJI Mavic is generally perceived as less annoying than the DJI Matrice.
    These observations provide valuable insights into the sound quality and annoyance levels associated with the tested drones and their respective audio signals.
\end{enumerate}

\begin{figure}[!t]
    \centering
    \includegraphics[width=1\linewidth]{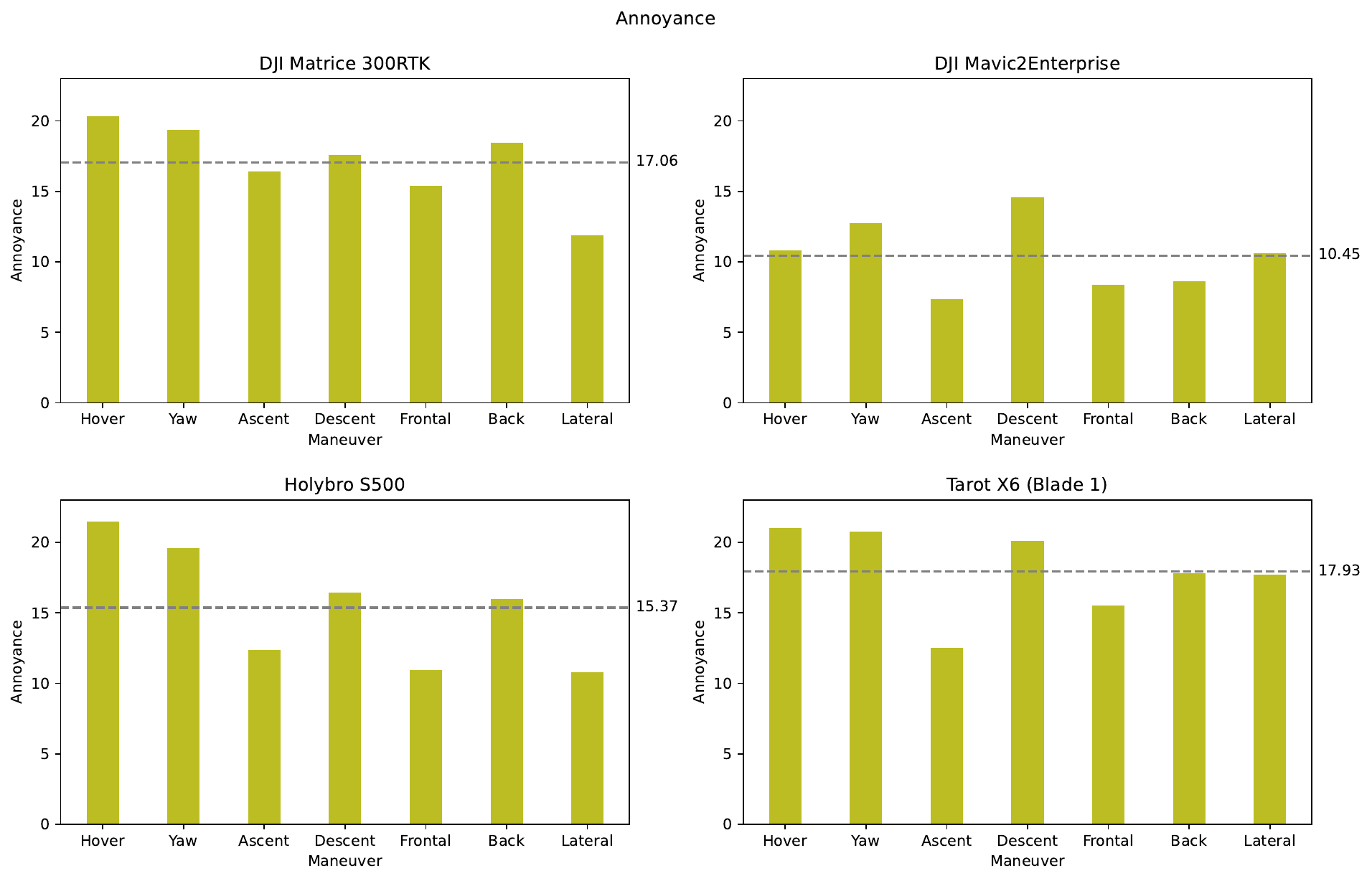}
    \caption{Comparison between annoyance perceived for each drone and maneuver}
    \label{fig:comparative_annoyance}
\end{figure}

\begin{table}[!b]
    \centering
    \begin{tabular}{|c|c|c|c|c|c|}
        \hline
        \rowcolor[HTML]{EFEFEF} 
        Drone        & PA    & L     & S    & R     & F     \\ \hline
        Tarot X6     & 17.93 & 21.22 & 2.11 & 0.14 & 0.02 \\ \hline
        DJI Matrice  & 17.06 & 20.02 & 1.87 & 0.27 & 0.01 \\ \hline
        Holybro S500 & 15.37 & 18.35 & 2.13 & 0.09 & 0.01 \\ \hline
        DJI Mavic    & 10.45 & 12.43 & 1.96 & 0.10 & 0.01 \\ \hline
    \end{tabular}
    \caption{SQL and PA values related to each drone}\label{tab:metrics_values}
\end{table}

\begin{figure}[!t]
    \centering
    \includegraphics[width=1\linewidth]{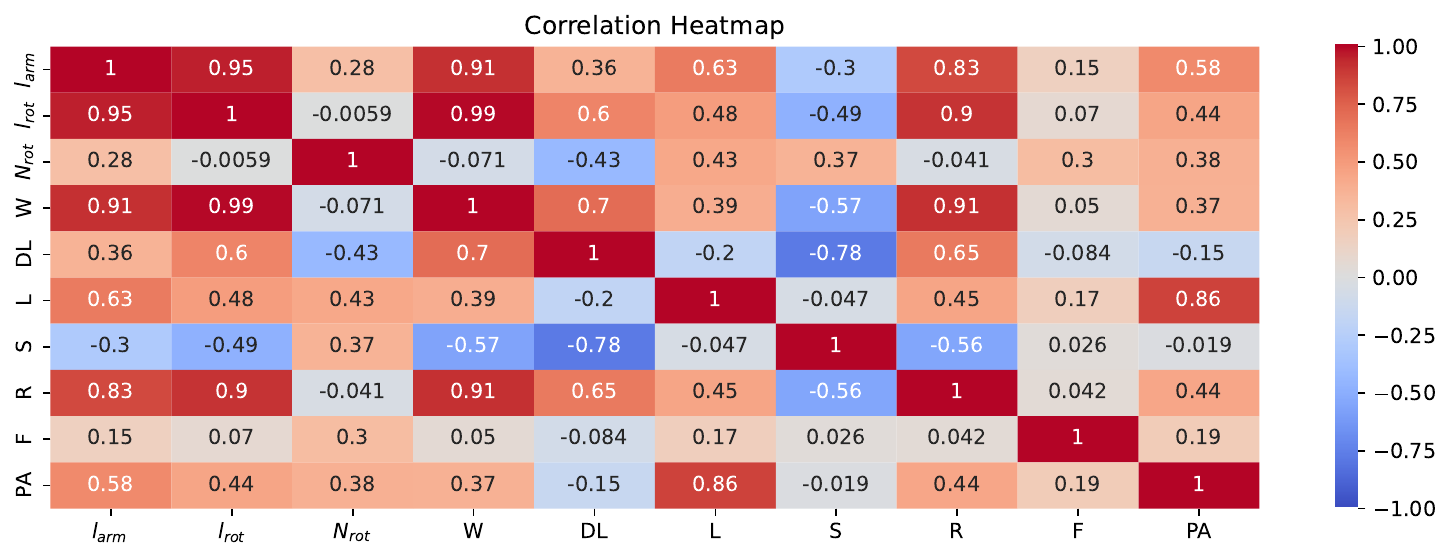}
    \caption{Heatmap of correlation between all drones, SQL and PA values}
    \label{fig:heatmap}
\end{figure}

From the average values of SQM and PA, average values for each maneuver were derived, as shown in Figure \ref{fig:comparative_annoyance} for PA. This approach allows for classifying the drones based on their level of "acoustic annoyance" generated. All index values are reported in Table \ref{tab:metrics_values}. The results indicate that the most annoying drone in the fleet is the Tarot X6, while the least annoying is the DJI Mavic 2 Enterprise. The data was collected in a dataset containing all index values for each maneuver and physical property. Additionally, Disc Loading is calculated from the available drone data. The Disc Loading is a parameter, often used in aerodynamics, that characterizes the distribution of lift-producing surfaces on an aircraft. It is defined as the total lift force generated by the aircraft (or drone) divided by the total area of its wings or rotors. In other words, it represents the amount of load supported per unit area of the lifting surface. Disc Loading can be calculated following the formula:
\[
DL = \frac{W}{N_{rot} \cdot \pi \cdot l_{rot}^2}
\]
Disc Loading was an interesting variable to measure because it provides insights into the efficiency and load distribution of the drone's lifting surfaces. Drones with a higher Disc Loading may have smaller wings or rotors relative to the total lift they generate, which could affect their flight characteristics and noise generation.

Finally, a heatmap, shown in Figure \ref{fig:heatmap}, was generated to determine the correlations between variables. This process allowed us to identify the variables most strongly associated with PA and to understand how PA is also correlated with all four SQM indices. By studying the values presented in the heatmap, important insights can be gained regarding the relationship between SQM, PA values and the physical parameters of the drone:
\begin{itemize}
    \item The values that most significantly influence PA are Loudness, with a correlation coefficient of 0.86, arm length $l_{arm}$, and rotor radius $l_{rot}$, along with Roughness.
    \item Loudness is also dependent on arm length $l_{arm}$ with a correlation value of 0.63, as well as rotor radius $l_{rot}$ and Roughness.
    \item Roughness remains primarily dependent on drone characteristics, such as weight, $l_{arm}$, and $l_{rot}$. It also exhibits a strong correlation with the Disc Loading value $DL$ and an equally strong inverse correlation with Sharpness.
    \item Sharpness shows an inverse correlation with several indices, with Disc Loading $DL$ being the most prominent, followed by $W$, Roughness, and $l_{rot}$. This means that high frequencies audio features could be assessed to drone's Disc Loading.
    \item Fluctuation strength does not appear to have significant correlations with the collected data.
    \item An important consideration is related to the value of $N_{rot}$ (number of rotors), which does not seem to participate in the found correlations. However, one would expect the number of rotors to influence the index values in some way. This limitation is likely due to the lack of other drones with a variable number of rotors, and the data's bias in this case does not allow for a good estimation of these correlations. Further experiments with other drones with different characteristics could improve the results of this study.
\end{itemize}
 
\section{Data prediction Methodology}\label{sec:model}
DL has revolutionized several fields, including computer vision, speech recognition, and text analysis, due to its remarkable success in classification and regression tasks~\cite{sarker2021deep}.
In this section, we describe the methodology used for the developing of a data--driven regression model for calculating psychoacoustic annoyance (PA) of a drone, based on a subset of features extracted from our dataset. The purpose of developing this model is to understand the relationships between PA and various physical parameters of the drones. 
\subsection{Feature Selection}\label{subsec:features}
The dataset analyzed in section \ref{sec:methods} was first processed to select the most relevant features for the prediction task. We created a Pandas DataFrame in which we derived the following subset of features, which are mostly independent of each other, according to the correlations shown in Figure \ref{fig:heatmap}:
\begin{enumerate}
    \item \textbf{Aspect Ratio (A/R):} Calculated as the ratio of arm length ($L_{arm}$) to rotor radius ($L_{rot}$). In this way we can obtain a single independent value.
    \item \textbf{Number of Rotors ($N_{rot}$):} Representing the drone's rotor configuration.
    \item \textbf{Weight ($W$):} Indicating the drone's mass.
    \item \textbf{Microphone's distance ($Mic_{dist}$):} Indicating the distance of the sound source to the microphone. All the distances were calculated from the hovering point, following the schema in Table \ref{tab:setup_data}.
    \item \textbf{Maneuver:} A categorical feature encoding different drone maneuvers. We converted it into numerical values based on a predefined mapping.
    \item \textbf{Psychoacoustic Annoyance (PA):} The target variable that we aim to predict.
\end{enumerate}
Feature selection process is important to remove or combine redundant or irrelevant attributes that could bring the model to be less accurate in general. The only two variables in the training set that remained somehow correlated are $A/R$ with $N_{rot}$, because keeping $N_{rot}$ resulted in a less overall loss of the model.
Additionally, we applied Min-Max scaling to normalize the data in order to ensure that the neural network can effectively learn from these features.

\subsection{Model architecture}
We developed and tested several AI and non-AI based architectures and tried to understand which model better fits our problem, i.e. Linear Regression, Support Vector Machine (SVM), Deep Neural Network (DNN), and Convolutional Neural Network (CNN) using both TensorFlow's Keras and scikit-learn Python API. Hyperparameter tuning was implemented using Nested Cross Validation (NCV). NCV estimates the generalization error of the underlying model and its hyperparameter search. The use of NCV is strongly advised in small datasets where test and train splitting can be noisy, indeed for methods that do not sufficiently control overfitting, such as K-fold cross-validation with small sample sizes, could produce biased performance estimates~\cite{vabalas2019machine}. 

\subsubsection{Custom Validation}
Due to the relatively small amount of data in the dataset, we developed a secondary validation test to assess which is the best model that has higher generalization capabilities. For this reason, we trained and optimized all four different models according to two main loss indices, namely RMSE (Full) and RMSE (Validation), which refers to the Root Mean Square Error (RMSE) in predicting the PA of a drone. This prediction is calculated as the mean of the predicted annoyance for each microphone (using the variable $Mic_{dist}$) and again averaged over all maneuvers of that drone. In the first case, the RMSE (Full) is the RMSE calculated from a model trained on the Training - Validation 0.2 splitted full dataset (with all examples available for all drones). RMSE (Validation), on the other hand, is calculated from four sub-models derived from each regression model (4 sub-models of 4 models for a total of 16 models). Specifically, for each unique drone in the dataset, a sub-training set is created and a new sub-model instance is trained without data related to the specific drone. A validation function is then used to evaluate the performance of the model trained on the sub-training set: this validation function first makes predictions on the data of the unseen drone, then, from all the predictions for each sub-model, calculates the mean PA for each unique drone in the dataset, and finally calculates the total RMSE error for all four predictions of all four sub-models. The results are shown in the table \ref{tab:modelcomparison}.

\begin{table}[]
    \centering
    \begin{tabular}{|c|c|c|}
        \hline
        \rowcolor[HTML]{EFEFEF} 
        Model  & RMSE (Full) & RMSE (Validation) \\ \hline
        DNN    & 0.3677      & 2.4313            \\ \hline
        CNN    & 1.1942      & 4.1761            \\ \hline
        SVM    & 0.2730      & 4.7847            \\ \hline
        Linear & 0.2474      & 7.0322            \\ \hline
    \end{tabular}
    \caption{Model performance comparison}
    \label{tab:modelcomparison}
\end{table}

As we can observe, although the RMSE calculated directly on all available data is significantly low due to the low amount of data and possible overfitting, the RMSE (Validation) provides a better understanding of the models' generalization capabilities. It is evident that the Linear Regressor model has the least amount of error; however, the other validation method illustrates its weak generalization capabilities due to the strong overfitting. This could be attributed to its inability to capture the non-linear characteristics of the data. Thus, we have decided to use the Deep Neural Network for our purpose, that is the model with the lower error on our RMSE validation test.

\subsubsection{Prediction model}

The model architecture consists of multiple densely connected layers, and it was designed to predict PA based on the selected features performing a regression task. The model architecture is designed as follows:
\begin{itemize}
    \item \textbf{Input Layer:} Comprising 5 neurons, corresponding to the selected features described in Section \ref{subsec:features}.
    \item \textbf{Hidden Layers:} 3 hidden layers, all utilizing the Rectified Linear Unit (ReLU) as activation function.
    \begin{itemize}
        \item Layer dense: 256 units
        \item Layer dense: 16 units
        \item Output layer: 1 unit
    \end{itemize}
    The dense architecture structure was based and optimized using Nested Cross Validation (NCV) for hyperparameter search, using different optimizer, layers, units and activation functions.
    \item \textbf{Output Layer:} A single neuron in the output layer, as this is a regression task aimed at predicting a the value (PA).
\end{itemize}
For our model compilation, we utilized the Adam optimizer, as suggested from the result of the hyperparameter research, and the mean squared error (MSE) loss function was implemented to calculate the model error. This specific loss function proves to be appropriate for regression tasks. 
\subsection{Model testing and results}
\begin{figure}[!t]%
    \centering
    \subfloat[\centering 
    ]{{\includegraphics[width=0.45\textwidth]{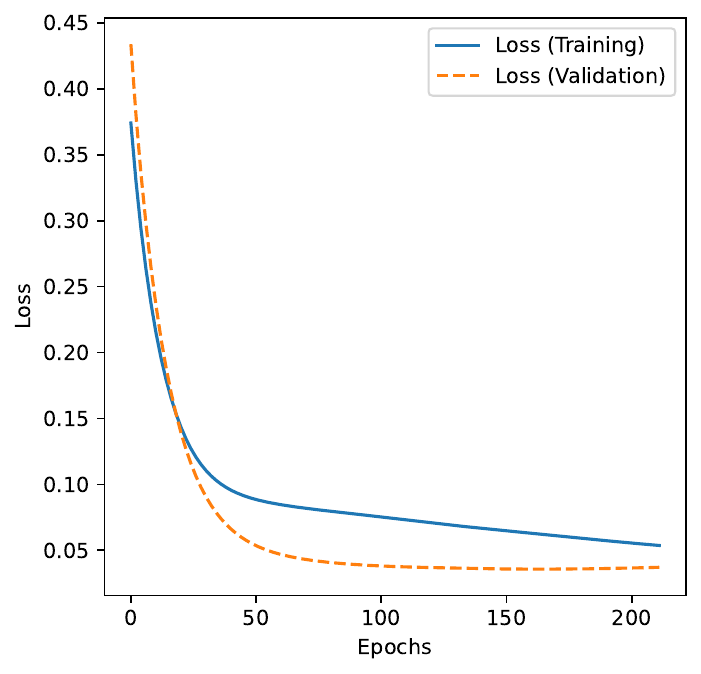} }}%
    \qquad
    \subfloat[\centering]
    {{\includegraphics[width=0.45\textwidth]{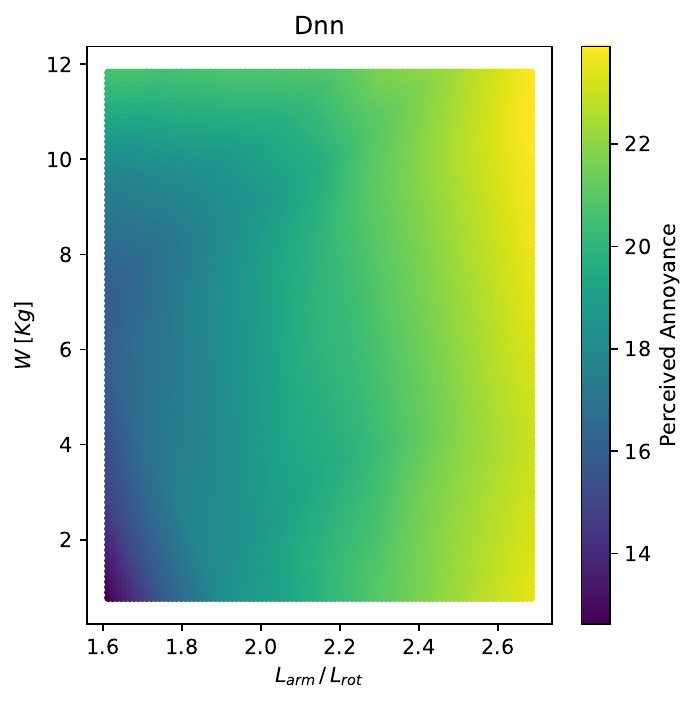} }}%
    \caption{(a) Plot of training history of the PA predictor model, (b) Visualizing PA predicted values varying $W$ and $L_{arm}/L_{rot}$}%
    \label{fig:example}%
\end{figure}

The training history of the model is an important indicator of its learning progress. We monitored the validation loss during training to assess the model's performance, and the history trend of the best model at the end of the NCV process is shown in Figure \ref{fig:example} (a). The model's predictions during the RMSE (Validation) phase, closely align with the ground truth values, indicating its ability to capture the relationships between the input features and PA. These results demonstrate the model's effectiveness in predicting PA. With this assumption, we can easily calculate and analyze the predicted correlation between drone's physical features and the PA. As we can see in Figure \ref{fig:example} (b), the increasing of Perceived Annoyance could be assessed to both overall drone's Weight and it's aspect ratio $L_{arm}/L_{rot}$. This result can be a starting point for a shape optimization process with the goal of building more efficient drones that produce less noise pollution.

\section{Conclusions}\label{sec:conclusions}
Our study extends current research by examining multiple drone models with multiple microphones in a realistic flying condition. This contributes to a better understanding of how design choices in drone manufacturing can affect noise generation. We conducted a thorough analysis of flight data collection and manipulation following the ISO standard, analyzing various flight maneuvers, drone physical characteristics, and facing the calculation and prediction of SQM and PA in real-world scenarios.

Our study can provide useful insights for the development of drone optimization strategies. There are several avenues for future research and improvement in this area. To further improve the predictive DL model, continuous refinement and expansion is necessary: incorporating data from additional drone models and a larger dataset will help to create a more robust and versatile model with better generalization capabilities.

In addition, the analysis can be improved by introducing new shape features for the drone. For example, incorporating rotor shape information can lead to a deeper understanding of how specific design elements affect acoustic characteristics. These additional features can improve our predictive model, resulting in more accurate and detailed assessments of perceived annoyance.

In summary, our study can represent a step forward in the ongoing effort to reduce the environmental and public acceptance impacts of drone noise. With continued research and refinement, there is potential to make drones quieter and more compatible with their surroundings.
\bibliography{biblio}

\end{document}